\documentclass[11pt]{article}
\usepackage{rotating}
\usepackage{amsbsy}

\begin{document}

       
\title{Calculation of reduced coefficients and matrix elements 
       in $jj$--coupling}

\author{ Gediminas Gaigalas$^{\, a,b}$ and Stephan Fritzsche$^{a}$
         \\
	 \\
	 \\
        $^a$ Fachbereich Physik, Universit\"a{}t Kassel,  \\
        Heinrich--Plett--Str. 40, D--34132 Kassel, Germany.   
	\\
        $^b$ Institute of Theoretical Physics and Astronomy,  \\
        A.\ Go\v{s}tauto 12, Vilnius 2600, Lithuania.
        \\ }

\maketitle

\date{}

\thispagestyle{empty}
\enlargethispage{0.0cm}

\begin{abstract}
A program RCFP will be presented for calculating standard quantities in the 
decomposition of many--electron matrix elements in atomic structure theory.
The list of quantities wich are supported by the present program includes 
the coefficients of fractional parentage, the reduced coefficients of 
fractional parentage, the reduced matrix elements of the unit operator 
$T^{(k)}$ as well as the completely reduced matrix elements of the operator 
$W^{(k_jk_q)}$ in $jj$--coupling. These quantities are now available for all
subshells ($nj$) with $j \leq 9/2$ including partially
filled $9/2$--shells. Our program is based on a recently developed  
new approach on the spin--angular integration which combines 
second quantization and quasispin methods with the theory of angular 
momentum in order to obtain a more efficient evaluation of many--electron 
matrix elements. An underlying Fortran 90/95 module can directly be 
used also in (other) atomic structure codes to accelerate the
computation for open--shell atoms and ions.
\end{abstract}

\newpage

{\large\bf PROGRAM SUMMARY}

\bigskip

{\it Title of program:} RCFP

\bigskip

{\it Catalogue number:} ADNA

\bigskip

{\it Program obtainable from:} CPC Program Library, 
     Queen's University of Belfast, N. Ireland. Users may obtain the
     program also by down--loading a tar--file \texttt{ratip-rcfp.tar} 
     from our home page at the University of Kassel    \newline
     (http://www.physik.uni-kassel.de/fritzsche/programs.html).

\bigskip

{\it Licensing provisions:} None.

\bigskip

{\it Computer for which the program is designed and has been tested:} \newline  
     IBM RS 6000, PC Pentium II.                            \newline
     {\it Installations:} University of Kassel (Germany).   \newline
     {\it Operating systems:} IBM AIX 4.1.2+, Linux 6.1+.
     
\bigskip

{\it Program language used in the new version:} ANSI standard Fortran 90/95.

\bigskip

{\it Memory required to execute with typical data:} 100 kB.

\bigskip

{\it No.\ of bits in a word:}  All real variables are parametrized by a 
     \texttt{selected kind parameter} and, thus, can be adapted to any 
     required precision if supported by the compiler. Currently, 
     the \texttt{kind} parameter is set to double precision 
     (two 32--bit words) as it is for other components of the RATIP package [1].

\bigskip

{\it Peripheral used:} Terminal for input/output.

\bigskip

{\it Distribution format:} tar gzip file

\bigskip

{\it CPC Program Library Subprograms required:}  
 Catalogue number: to be assigned:
 Title: REOS99;   Ref. [1]          \\


\bigskip


\bigskip

{\it Nature of the physical problem:}  \newline
     The calculation of atomic properties and level structures is based on the
     evaluation of many--particle matrix elements of physical operators.
     For symmetry--adapted functions, the matrix element for a given
     tensor operator $A^{(K)}$ of rank $K$  can be expressed as  
     \newline
     $\sum_{j,k}~{\rm coeff} (j,k)~<\gamma_{j}J_{j}||A^{(K)}||\gamma_{k}J_{k}>$
     by using the (reduced) coefficients of fractional parentage and the
     reduced matrix elements of the (unit) standard tensors 
     $T^{(k)}$ or $W^{(k_q k_j)}$. 
     These reduced coefficients and matrix elements are
     frequently applied to both the configuration interaction and
     multi--configuration Dirac--Fock method~[2] as well as to many--body
     perturbation theory [3].
          
\bigskip

{\it Method of solution:}  \newline
     A new combination of second quantization and quasispin methods
     with the theory of angular momentum and irreducible tensor operators
     leads to a more efficient evaluation of (many--particle) matrix elements
     and to faster computer codes~[4]. Practical implementations of this new
     scheme will support not only large--scale
     computations on open--shell atoms but may even help to develop programs
     for calculating the angular parts of (effective) one-- and two--particle
     operators for many--body perturbation theory (in higher orders) in the
     future.

\bigskip

{\it Restrictions onto the complexity of the problem:}  \newline
     For $jj$--coupled subshells states, our module provides coefficients 
     and matrix elements
     for all subshells ($nj$) with $j$ = $1/2$, $3/2$, $5/2$, $7/2$, and $9/2$.

\bigskip

{\it Typical running time:}  \newline
     For large computations, the running time depends on the shell 
     structure and the size of the wave function expansion for a 
     given atomic system. However, the program \textit{promptly} 
     responds in its interactive mode if only single
     coefficients and matrix elements are to be calculated.

\bigskip

{\it Unusual features of the program:} \newline
     The interactive version of RCFP can be used as an
     ''electronic tabulation'' of standard quantities for evaluating 
     general matrix elements for $jj$--coupled functions.

\bigskip

{\it References:}   \newline
     [1] S.\ Fritzsche, C.\ F.\ Fischer, and C.\ Z.\ Dong, 
         Comput.\ Phys.\ Commun.\ 124 (2000) 340.
     \newline
     [2] I.\ P.\ Grant, and H.\ Quiney, Advances in Atomic and Molecular
         Physics 23 (1987) 37.
     \newline
     [3] G.\ Merkelis, G.\ Gaigalas, J.\ Kaniauskas, and Z.\ Rudzikas,
         Izvest.\ Acad.\ Nauk SSSR, \newline
     \hspace*{0.55cm}Phys.\ Series 50 (1986) 1403.
     \newline
     [4] G.\ Gaigalas, Lithuanian Journal of Physics 39 (1999) 80.
         
\newpage
{\large\bf LONG WRITE--UP}

\bigskip

\section{Introduction}

In atomic structure theory, the efficient evaluation of many--electron
matrix elements play a crucial role. Typically, such matrix elements have to
be evaluated for different one-- and two--particle operators which describe
the interaction of the electrons with each other or with external
particles and fields. By exploiting the techniques of Racah's 
algebra~\cite{RI}, the evaluation of these matrix elements may often be
considerably simplified by carrying out the integration over 
the spin--angular coordinates analytically.
For atoms with open shells, several approaches to this analytic integration
were developed in the past. One of the most widely--used computational
schemes is from Fano \cite{Fano,Grant-a} and has been implemented in a 
number of powerful programs~\cite{MCHF-ASP,MGN} since that time.

Fano's procedure \cite{Fano} is based on the coefficients of fractional 
parentage (cfp). During the last decades, this procedure
was applied both to $LS$-- and $jj$--coupling;
in the following, we will restrict ourselves to $jj$--coupling 
as appropriate for relativistic calculations.
By using the cfp as the basic quantities, however, Fano's procedure 
does not exploit the full power of Racah's algebra. Instead of using cfp,
it is often more convenient to exploit unit tensors \cite{Judd-o,R}
which are closer related to the tensorial
form of physical operators. But so far, unit tensors have been
applied only for evaluating diagonal matrix elments while all non--diagonal 
matrix elements still have to be evaluated by using the cfp \cite{R,NR}. 
A recently developed approach \cite{method2,method6} now
treats diagonal and non--diagonal matrix elements on a uniform basis.
This approach is based on a second quantization and uses a coupled 
tensorial form for the creation and annihilation operators \cite{RK-book}. 
It also  applies the theory of angular momentum to two different spaces, 
i.e. the space of orbital angular momentum $j$ and the quasispin space
\cite{RK}. The basic quantities of this new approach are the so--called reduced 
coefficients of fractional parentage (rcfp) and the completely 
reduced matrix elements of the $W^{(k_qk_j)}$ operator.

Obviously, each computational scheme is based on a set of standard quantities
to decompose the many--electron matrix elements.
These quantities are either the cfp, rcfp, the reduced matrix elements
of the unit tensor $T^{(k)}$, the completely reduced matrix elements 
$W^{(k_qk_j)}$, depending on the approach.
Therefore, very different tabulations of these quantities are found in the 
literature. For example, numerical values for the cfp are found
by Shalit and Talmi~\cite{ST} for subshells with $j=5/2$, $7/2$ and 
$9/2$ while rcfp  for $j=5/2$ and $7/2$ were first tabulated by 
Savi\v cius {\it et al}~\cite{SKR}. Matrix elements 
of $T^{(k)}$ are tabulated, for instance, by Slepcov {\it et al}~\cite{SSKR} 
for subshells with $j=3/2$, $5/2$ and $7/2$; often, however, it is more 
convenient to express these matrix elements in terms of the completely
reduced matrix elements of the operator $W^{(k_q k_j)}$ even though no explicit 
compilation of these matrix elements yet available. In practical 
applications, they are instead derived from a sum of products of rcfp and 
6--$j$ symbols.

In this paper, we will present the program RCFP for the calculation of
the standard quantities both in Fano's and our new approach. 
These quantities are needed for the integration over the 
spin--angular variables. Our program not only supports large--scale
computations on open--shell atoms but may even help to develope codes
for calculating the angular parts of (effective) one-- and two--particle
operators from many--body perturbation theory (in higher orders) in the
future.

The theoretical background will be presented in section 2. This includes a brief
outline of the quasispin concept, the definitions of the rcfp and
the reduced matrix elements of the unit tensors $W^{(k_q k_j)}$  and $T^{(k)}$ 
as well
as the proper classification of subshells in $jj$--coupling.
The program organization will be dealt with in section 3 while, finally,
a few examples are given in section 4.

\section{Theoretical background}

The theory of angular integration for symmetry--adapted functions
has been reviewed in several texts and monographs 
\cite{Grant-a,R,method6,Cowan}. As mention above, this theory is usually 
built on a number of standard quantities like the cfp
or the reduced matrix elements of the unit tensor $T^{(k)}$ which, in turn,
can be applied to lay down the expressions for 
more complex matrix elements. Other important quantities 
(which are also provided by our program) are the rcfp and the completely 
reduced matrix elements of the tensor operators $W^{(k_q k_j)}$. In 
the following, we shall not repeat too many details about this angular 
integration on the spin--angular coordinates; instead, 
we just list the definition of those quantities which can be obtained from 
our program  along with a number of useful relations among them. 
For all further details, we ask the reader to refer to the literature 
given above.

\medskip

In the literature, several definitions and phase conventions are used for 
defining the standard quantities for angular integration. Here, we follow 
the definitions from Savi\v cius~\cite{SKR} and from Kaniauskas and 
Rudzikas~\cite{RK}. We also apply the so--called 
\textit{standard--phase systems},
\begin{equation}
   \label{eq:phase}
   A_{m_k}^{\left(k \right) \dagger} = \left( -1 \right) ^{k-m_k}
   A_{-m_k}^{\left(k \right)}
\end{equation}
  
throughout this paper which were originally introduced by Fano and 
Racah~\cite{FR-phase}.

\subsection{The quasispin concept}

In $jj-$coupling, a wave function for a subshell of $N$ 
\textit{equivalent} electrons with principal quantum number $n$ and
(one--electron) angular momentum $j$ is often written as
\begin{equation}
   \label{eq:quasispin-aa}
   |nj^N\alpha J)
\end{equation}
where $J$ denotes the total angular momentum and $\alpha$ all additional
quantum numbers as needed for an unique classification of these states.
Using the quasispin concept, a further (angular) quantum number
$Q$, the quasispin momentum of the subshell, can be introduced so that the
wave function of this subshell (to which we briefly refer to as a
 \textit{subshell state}) then reads
\begin{equation}
   \label{eq:quasispin-ab}
   |nj^N\alpha QJ) \ .
\end{equation}
For any given subshell, the quasispin Q is closely related to the seniority
quantum number $\nu$ as used in the 
\textit{seniority scheme}, i.e.\  $Q=\left( \frac{2j+1}{2}-\nu \right) /2$.
If compared with the seniority notation the quasispin $Q$ to has the advantage
of its projection, $M_Q$, being related to the
occupation number $N$ by $M_Q=\left( N - \frac{2j+1}{2} \right) /2$. 
Therefore, when exploring the quasispin concept for classifying the atomic 
subshell states (\ref{eq:quasispin-ab}), the Wigner--Eckart theorem can 
be applied 
twice, both to the space of quasispin ($Q$--space) as well as to the total 
angular momentum ($J$--space). Hence, any reduced matrix element in $J$--space 
can be further reduced also in $Q$--space \cite{RK}
\begin{eqnarray}
   \label{eq:quasispin-a}
   \left( j^N\;\alpha QJM_Q||\, A_{m_q}^{\left( qj\right) } \,
                           ||j^{N^{\prime }}\;
   \alpha ^{\prime} Q^{\prime } J^{\prime } M_Q^{\prime } \right)   
   & = & (-1)^{Q-M_Q} \,
   \left(
   \begin{array}{ccc}
   Q & q & Q^{\prime } \\
   -M_Q & m_q & M_Q^{\prime }
   \end{array}     
   \right)  
   \nonumber  \\[0.3cm]
   &  & \hspace*{1.5cm}
   \times \,
   \left( j\;\alpha QJ|||\, A^{\left( qj\right) } \,|||j\;
   \alpha ^{\prime} Q^{\prime } J^{\prime} \right)
\end{eqnarray}
to a so--called \textit{completely reduced} matrix element. 
In Eq.\ (\ref{eq:quasispin-a}), $ A_{\,m_q}^{\left( qj\right)} $ denotes 
a tensor with rank $q$ and projection $m_q$ in the $Q-$space. 
As seen from its notation the completely reduced matrix element 
$\left( j\;\alpha QJ|||\, A^{\left( qj\right) } \,|||j\;\alpha ^{\prime
}Q^{\prime }J^{\prime }\right)$
is independent of the occupation number $N$ of the particular subshell
states; the occupation number $N$ of these states occurs explicitly only 
on the left--hand--side 
of Eq.\ (\ref{eq:quasispin-a}) while it is incorporated into $Q$ on the
right--hand side. Thus, by applying the quasispin concept, the evaluation 
of general matrix elements will result in a much smaller number of
completely reduced matrix elements which are independent of the occupation of
electrons $N$ in this subshell but still allows an unique decomposition.

\subsection{Coefficients of fractional parentage}

The electron creation $ a_{m_{j}}^{\left(j \right)} $ and annihilation 
$ a_{-m_{j}}^{\left(j \right) \dagger} $ operators play a key role in 
the theory of second quantization and atomic structure \cite{Judd-s}. 
Using the quasispin concept, the 
operators $a_{-m_{j}}^{\left(j \right)}$ and 
$ \stackrel{\sim}{a}_{m_{j}}^{\left(j \right)}=\left( -1\right)^{j-m_{j}}
  a_{-m_{j}}^{\left(j \right) \dagger}$ also form components of an 
irreducible tensor of rank $q=\frac 12$ in $Q$--space, i.e.\
\begin{equation}
   \label{eq:second-a}
   a_{\, m_q m_j }^{\left( q j\right) }=\left\{
   \begin{array}{ll}
   a_{\, m_j}^{( j )}  & \hspace*{1cm} \mbox{ for } m_q  = \frac 12, \\[0.3cm]
   \stackrel{\sim }{a}_{\, m_j}^{( j ) } 
                       & \hspace*{1cm} \mbox{ for } m_q = -\frac 12.
   \end{array}
   \right.  
\end{equation}
Compared with the electron creation and annihilation operators above, 
the operators $a_{\, m_q m_j }^{\left( q j\right) }$ also act in an additional 
quasispin space like a tensor component with
rank $q$ and a projection $m_q=\pm \frac 12$.
There is the following relation known between the reduced matrix element 
of a creation operator and the cfp \cite{La-Ma}
\begin{eqnarray}
   \label{eq:cfp-amat}
   \left( j^N \;\alpha QJ||a^{\left( j \right) }||j^{N-1} \;\alpha ^{\prime
   }Q^{\prime }J^{\prime }\right)
   & = &
   \left( -1\right) ^{N}
   \sqrt{N\left[ J\right]}
   \left( j^N\;\alpha QJ||j^{N-1}\;
   \left( \alpha ^{\prime }Q^{\prime}J^{\prime }
   \right) j
   \right)
\end{eqnarray}
where $\left[ J \right] \equiv \left( 2J +1\right)$.
Eq.\ (\ref{eq:cfp-amat}) can be used to define the relation between the
cfp and its reduced counterpart in $Q$--space. Introducing the $z-$projection, 
$M_Q$, of the quasispin, this relation is given by \cite{R}
\newpage
\begin{eqnarray}
   \label{eq:second-b}
   \left( j\;\alpha QJ|||a^{\left( qj\right) }|||j\;\alpha ^{\prime
   }Q^{\prime }J^{\prime }\right)
   & = &
   \left( -1\right) ^{N+Q-M_Q}
   \sqrt{N\left[ J\right]}  \, \left(
   \begin{array}{ccc}
   Q & 1/2 & Q^{\prime } \\
   -M_Q & 1/2 & M_Q^{\prime }
   \end{array}  \right)^{-1} 
   \nonumber \\[0.3cm]
   &  & \hspace*{2.7cm} \times
   \left( j^N\;\alpha QJ||j^{N-1}\;
   \left( \alpha ^{\prime }Q^{\prime
   }J^{\prime }
   \right) j
   \right).
\end{eqnarray}
The properties of the rcfp 
have been summarized by Savi\v cius {\it et al}~\cite{SKR} and Gaigalas 
{\it et al.}~\cite{method5}. The latter reference also discusses
\textit{phase conventions} which are frequently applied in the literature 
to subshell states with a the same number $N \; (< j + 1/2) \,$ of electrons 
or holes, respectively. 

\subsection{Reduced matrix elements of standard operators}

The unit tensors $W^{(k_qk_j)}$ and $T^{(k)}$ are other standard quantities in
atomic spectroscopy. Many texts on the evaluation of matrix elements in
many--particle physics frequently refer to these quantities 
\cite{Judd-o,R}. The tensor $W^{(k_qk_j)}$,
for example, is defined as the tensorial product of two creation
operators in second quantization
\begin{equation}
   \label{eq:unit-a}
   W^{\, (k_q k_j) }_{\, m_q m_j} \: = \:
   \left[ \,  a^{ ( q j ) } \times a^{ ( q j ) } 
   \right]^{\, (k_q k_j) }_{\, m_q m_j}.
\end{equation} 

Following Savi\v cius {\it et al}~\cite{SKR}, the operators $T^{(k)}$ and 
$W^{(k_qk_j)}$ obey the relation
\begin{equation}
   \label{eq:unit-d}
   T_{\,m}^{\,(k) } = \left\{
   \begin{array}{ll}
     - \left( 2 \left[ k \right] \right) ^{-1/2} W_{\,0m}^{\, (0 k) }  
       & \hspace*{1.5cm} \mbox{if $k$ is odd}, \\[0.35cm]
     - \left( 2 \left[ k \right] \right) ^{-1/2} W_{\,0m}^{\, (1 k) }  
       & \hspace*{1.5cm} \mbox{if $k$ is even }.
   \end{array}
   \right.
\end{equation}

The reduced matrix elements of $T^{(k)}$ can be represented in terms of 
a sum over 6--$j$ symbols and cfp's
\begin{eqnarray}
   \label{eq:unit-e}
   \hspace*{-0.5cm}
   \left( j^N\;\alpha J || T^{(k)} || j^{N}\; 
         \alpha ^{\prime } J^{\prime }\right) 
   & = & 
   N \sqrt{ \left[ J,J^{\prime} \right]} \;
   \displaystyle {\sum_{\alpha ^{\prime \prime } J^{\prime \prime }}}
                   (-1)^{J^{\prime \prime } +j + J + k} \,
   \left\{
   \begin{array}{ccc}
      j           & J & J^{\prime \prime } \\
      J^{\prime } & j & k
   \end{array} 
   \right\} \hspace*{0.6cm}
   \nonumber \\[0.35cm]
   &  & \hspace*{0.3cm} \times
   \left( j^N\;\alpha J|| j^{N-1}\;\left( \alpha ^{\prime \prime }
          J^{\prime \prime } \right) j\right)
   \left( j^{N-1}\;\left( \alpha ^{\prime \prime } 
          J^{\prime \prime } \right) j 
          || j^N\;\alpha ^{\prime } J^{\prime } \right) \; .
\end{eqnarray}
The completely reduced matrix elements of the operator $W^{(k_qk_j)}$ is related
to the rcfp in the following way
\begin{eqnarray}
   \label{eq:unit-f}
   &  & \hspace*{-1.6cm}
   \left( nj\;\alpha QJ||| W^{(k_qk_j)} |||nj\;\alpha
          ^{\prime }Q^{\prime }J^{\prime }\right) 
   \nonumber \\[0.3cm]
   & = &
   \left( -1\right) ^{Q+J+Q^{\prime }+J^{\prime }
          +k_q+k_j} \, \sqrt{\left[ k_q,k_j \right]}  \;
   \displaystyle {\sum_{\alpha ^{\prime \prime }Q^{\prime \prime }J^{\prime
                  \prime }}} \;
   \left\{
   \begin{array}{ccc}
      q & q & k_q \\
      Q^{\prime } & Q & Q^{\prime \prime }
   \end{array}
   \right\} \left\{
   \begin{array}{ccc}
      j & j & k_j \\
      J^{\prime } & J & J^{\prime \prime }
   \end{array}
   \right\}  
   \nonumber  \\[0.4cm]
   &  & \hspace*{0.8cm} \times \,                 
   \left( j\;\alpha QJ|||a^{\left( qj\right)} ||| 
          j\;\alpha ^{\prime \prime }Q^{\prime \prime }J^{\prime \prime }
   \right)  
   \left( j\;\alpha ^{\prime \prime }Q^{\prime \prime }J^{\prime \prime } |||
   a^{\left( qj\right)} ||| 
   j \;\alpha ^{\prime }Q^{\prime }J^{\prime }\right) \, .
\end{eqnarray}
Thus, a close relationship between the completely reduced matrix elements 
of $W^{(k_qk_j)}$  and the reduced matrix elements of the unit tensor 
$T^{(k)}$ 
is given by
\begin{eqnarray}
   \label{eq:unit-g}
   \left( nj \;\alpha QJ ||| W^{(1k)} |||
          nj \;\alpha ^{\prime }Q^{\prime }J^{\prime }\right) 
   & = &
   (-1)^{1+Q-M_Q} \sqrt{2 \left[ k \right] }
   \left(
   \begin{array}{ccc}
     Q   & 1 & Q^{\prime } \\
     -M_Q & 0 & M_Q^{\prime }
   \end{array}
   \right)^{-1} 
   \nonumber  \\[0.2cm]
   &  & \hspace*{1.0cm} \times \,
   \left( j^N\;\alpha QJ M_Q || T^{(k)} || j^{N}\; \alpha ^{\prime } 
   Q^{\prime } J^{\prime } M_{Q}^{\prime } \right)
   \nonumber \\[0.4cm] 
   &  & \hspace*{4cm} \mbox{if $k_q$ = 1 and $k$ is even}
\end{eqnarray}
and
\begin{eqnarray}
   \label{eq:unit-h}
   \left( nj \;\alpha QJ ||| W^{(0k)} |||
          nj \;\alpha ^{\prime }Q^{\prime }J^{\prime }\right)
   & = &
   - \sqrt{2 \left[ Q,k \right] } \,
   \left( j^N\;\alpha QJ M_Q || T^{(k)} || j^{N}\; \alpha ^{\prime } Q^{\prime } 
          J^{\prime } M_{Q}^{\prime } \right)
   \nonumber \\[0.4cm] 
   &  & \hspace*{4cm} \mbox{if $k_q$ = 0 and $k$ is odd.}
\end{eqnarray}
Since the completely reduced matrix elements 
$\left( nj\;\alpha QJ|||\, W^{ (k_q k_j) } \,|||nj\;\alpha
       ^{\prime }Q^{\prime }J^{\prime }\right) $ of the operator 
$W^{(k_qk_j)}$ are, again, independent of the occupation number, they 
allow for a more compact representation (tabulation) in atomic structure
calculations. This fact becomes important, in particular, when calculating
atoms with open $d-$ and/or $f-$shells. So far, no detailed analysis 
or tabulation of these completely reduced matrix elements in $jj-$coupling
has been published in the literature or has been implemented in any 
atomic structure code.

\subsection{Classification of subshells in $jj$--coupling}

A unique classification of the atomic states and, hence, the subshell states
is required for all structure computations. For subshells with 
$j = 1/2, \, 3/2, \, 5/2, $ and $ 7/2 $, two quantum numbers 
$Q$ and $J$ (respectively, $\nu$ and $J$ in the
seniority notation) are sufficient to classify the subshell states for all
allowed occupation numbers $N$ unambiguously.
For these subshells, no additonal quantum numbers $\alpha$ are then needed 
to be specified in (\ref{eq:quasispin-ab}). By contrast, some additional
number(s) are required for classifying the subshell states for
$j \ge 9/2$ [cf.\ Shalit and Talmi~\cite{ST} or 
Grant~\cite{Grant-a}]. For $j = 9/2$, there are two doublets 
(pairs of subshell states) with $\nu = 4, \: J = 4$ and $\nu = 4, \: J = 6$ 
in the $[9/2]^4$ and $[9/2]^6$ configurations which require an additional 
''number'' in order to classify these states uniquely.
To distinguish the individual subshell states of these two pairs, we
use the number $Nr = 1$ or $Nr = 2$ beside of the standard
quantum numbers $Q$ and $J$, respectively, $\nu$ and $J$.
Table~1 lists all ($jj-$coupled) subshell states 
for $j$ = $1/2$, $3/2$, $5/2$, $7/2$ and $9/2$, starting for each $j$ 
with the lowest occupation number.

\begin{small}
\begin{table}
\vspace*{-1.0cm}
{\bf Table 1}
\hspace{0.2cm}
{\rm Allowed couplings $\left[ j \right]^{N}$ of $N$ \textit{equivalent}
electrons for subshells
with $j = 1/2, \, \ldots, \, 9/2$. The seniority quantum number 
$\nu$, the subshell angular momentum $J$, the subshell quasispin $Q$ and 
the number $Nr$ (for subshells with $j = 9/2$ only) are shown.}

\begin{footnotesize}
\begin{center}
\begin{tabular}{ l l r l l|l l r l l }
\hline \hline  \\
$subshell$ & $\nu$ & $J$ & $2Q$ & $Nr$ &
$subshell$ & $\nu$ & $J$ & $2Q$ & $Nr$  \\ \hline
           &       &     &      &      &
           &       &     &      &       \\
$\left[ 1/2 \right]^{0}$ or $\left[ 1/2 \right]^{2}$ & $0$ & $ 0 $ & $1$ &   &
                                                     & $3$ & $5/2$ & $2$ &   \\
$\left[ 1/2 \right]^{1}$                             & $1$ & $1/2$ & $0$ &   &
                                                     & $3$ & $7/2$ & $2$ &   \\
                                                     &     &       &     &   &
                                                     & $3$ & $9/2$ & $2$ &   \\
$\left[ 3/2 \right]^{0}$ or $\left[ 3/2 \right]^{4}$ & $0$ & $ 0 $ & $2$ &   &
                                                     & $3$ & $11/2$& $2$ &   \\
$\left[ 3/2 \right]^{1}$ or $\left[ 3/2 \right]^{3}$ & $1$ & $3/2$ & $1$ &   &
                                                     & $3$ & $13/2$& $2$ &   \\ 
$\left[ 3/2 \right]^{2}$                             & $0$ & $ 0 $ & $2$ &   &
                                                     & $3$ & $15/2$& $2$ &   \\ 
                                                     & $2$ & $ 2 $ & $0$ &   &
                                                     & $3$ & $17/2$& $2$ &   \\ 
                                                     &     &       &     &   &
                                                     & $3$ & $21/2$& $2$ &   \\
$\left[ 5/2 \right]^{0}$ or $\left[ 5/2 \right]^{6}$ & $0$ & $ 0 $ & $3$ &   &
$\left[ 9/2 \right]^{4}$ or $\left[ 9/2 \right]^{6}$ & $0$ & $ 0 $ & $5$ &   \\	
$\left[ 5/2 \right]^{1}$ or $\left[ 5/2 \right]^{5}$ & $1$ & $5/2$ & $2$ &   &
                                                     & $2$ & $ 2 $ & $3$ &   \\
$\left[ 5/2 \right]^{2}$ or $\left[ 5/2 \right]^{4}$ & $0$ & $ 0 $ & $3$ &   &
                                                     & $2$ & $ 4 $ & $3$ &   \\
                                                     & $2$ & $ 2 $ & $1$ &   &
                                                     & $2$ & $ 6 $ & $3$ &   \\
                                                     & $2$ & $ 4 $ & $1$ &   &
                                                     & $2$ & $ 8 $ & $3$ &   \\
$\left[ 5/2 \right]^{3}$                             & $1$ & $5/2$ & $2$ &   &
                                                     & $4$ & $ 0 $ & $1$ &   \\
                                                     & $3$ & $3/2$ & $0$ &   &
                                                     & $4$ & $ 2 $ & $1$ &   \\
                                                     & $3$ & $9/2$ & $0$ &   &
                                                     & $4$ & $ 3 $ & $1$ &   \\
                                                     &       &     &     &   &
                                                     & $4$ & $ 4 $ & $1$ & 1 \\
$\left[ 7/2 \right]^{0}$ or $\left[ 7/2 \right]^{8}$ & $0$ & $ 0 $ & $4$ &   &
                                                     & $4$ & $ 4 $ & $1$ & 2 \\
$\left[ 7/2 \right]^{1}$ or $\left[ 7/2 \right]^{7}$ & $1$ & $7/2$ & $3$ &   &
                                                     & $4$ & $ 5 $ & $1$ &   \\
$\left[ 7/2 \right]^{2}$ or $\left[ 7/2 \right]^{6}$ & $0$ & $ 0 $ & $4$ &   &
                                                     & $4$ & $ 6 $ & $1$ & 1 \\
                                                     & $2$ & $ 2 $ & $2$ &   &
                                                     & $4$ & $ 6 $ & $1$ & 2 \\
                                                     & $2$ & $ 4 $ & $2$ &   &
                                                     & $4$ & $ 7 $ & $1$ &   \\
                                                     & $2$ & $ 6 $ & $2$ &   &
                                                     & $4$ & $ 8 $ & $1$ &   \\
$\left[ 7/2 \right]^{3}$ or $\left[ 7/2 \right]^{5}$ & $1$ & $7/2$ & $3$ &   &
                                                     & $4$ & $ 9 $ & $1$ &   \\
                                                     & $3$ & $3/2$ & $1$ &   &
                                                     & $4$ & $10 $ & $1$ &   \\
                                                     & $3$ & $5/2$ & $1$ &   &
                                                     & $4$ & $12 $ & $1$ &   \\
                                                     & $3$ & $9/2$ & $1$ &   &
$\left[ 9/2 \right]^{5}$                             & $1$ & $9/2$ & $4$ &   \\
                                                     & $3$ & $11/2$& $1$ &   &
                                                     & $3$ & $3/2$ & $2$ &   \\
                                                     & $3$ & $15/2$& $1$ &   &
                                                     & $3$ & $5/2$ & $2$ &   \\
$\left[ 7/2 \right]^{4}$                             & $0$ & $ 0 $ & $4$ &   &
                                                     & $3$ & $7/2$ & $2$ &   \\
                                                     & $2$ & $ 2 $ & $2$ &   &
                                                     & $3$ & $9/2$ & $2$ &   \\
                                                     & $2$ & $ 4 $ & $2$ &   &
                                                     & $3$ & $11/2$& $2$ &   \\
                                                     & $2$ & $ 6 $ & $2$ &   &
                                                     & $3$ & $13/2$& $2$ &   \\
                                                     & $4$ & $ 2 $ & $0$ &   &
                                                     & $3$ & $15/2$& $2$ &   \\
                                                     & $4$ & $ 4 $ & $0$ &   &
                                                     & $3$ & $17/2$& $2$ &   \\
                                                     & $4$ & $ 5 $ & $0$ &   &
                                                     & $3$ & $21/2$& $2$ &   \\
                                                     & $4$ & $ 8 $ & $0$ &   &
                                                     & $5$ & $1/2$ & $0$ &   \\
                                                     &       &     &     &   &
                                                     & $5$ & $5/2$ & $0$ &   \\
$\left[ 9/2 \right]^{0}$ or $\left[ 9/2 \right]^{10}$& $0$ & $ 0 $ & $5$ &   &
                                                     & $5$ & $7/2$ & $0$ &   \\
$\left[ 9/2 \right]^{1}$ or $\left[ 9/2 \right]^{9}$ & $1$ & $9/2$ & $4$ &   &
                                                     & $5$ & $9/2$ & $0$ &   \\
$\left[ 9/2 \right]^{2}$ or $\left[ 9/2 \right]^{8}$ & $0$ & $ 0 $ & $5$ &   &
                                                     & $5$ & $11/2$& $0$ &   \\
                                                     & $2$ & $ 2 $ & $3$ &   &
                                                     & $5$ & $13/2$& $0$ &   \\
                                                     & $2$ & $ 4 $ & $3$ &   &
                                                     & $5$ & $15/2$& $0$ &   \\
                                                     & $2$ & $ 6 $ & $3$ &   &
                                                     & $5$ & $17/2$& $0$ &   \\
                                                     & $2$ & $ 8 $ & $3$ &   &
                                                     & $5$ & $19/2$& $0$ &   \\
$\left[ 9/2 \right]^{3}$ or $\left[ 9/2 \right]^{7}$ & $1$ & $9/2$ & $4$ &   &
                                                     & $5$ & $25/2$& $0$ &   \\
                                                     & $3$ & $3/2$ & $2$ &   \\
\hline
\end{tabular}
\end{center}
\end{footnotesize}
\end{table}
\end{small}

\section{Program organization}

\subsection{Overview to program}

The program RCFP supports the computation of the cfp,
the rcfp, the (completely)
reduced matrix elements of the operator $W^{(k_qk_j)}$ as well as  
the matrix elements of the unit tensor $T^{(k)}$. It can be applied 
interactively, for instance, for calculating a few individual coefficients 
or matrix elements  in some theoretical derivation but also, by \textit{using}
the underlying module \texttt{rabs\_{}rcfp}, in any relativistic atomic 
structure
calculations in order to evaluate all required (many--electron) matrix 
elements automatically. RCFP is written in Fortran 90/95 and is designed 
as additional component of the RATIP package \cite{Fritzsche/CFF/Dong:99}
as will be explained in subsection 3.3. By exploiting the advantages 
of the new Fortran 90/95 standard, 
we defined several derived data types which facilitate the work and which 
shall enable us to incorporate this module in our present developments 
on large--scale computations for open--shell atoms and ions. 
The definition of the various derived structures can be found in the header 
of the module \texttt{rabs\_{}rcfp} but will not be explained here.

\medskip

As seen from section 2, the most basic quantities for evaluating 
matrix elements among different subshell states are the rcfp and 
the completely
reduced matrix elements of $W^{(k_q k_j)}$. These quantities are more general
than the cfp or the reduced matrix elements of the unit tensor
$T^{(k)}$ as they do not depend on the occupation number in the corresponding
shells. Thus, the rcfp and the completely reduced matrix elements can be
tabulated much easier for subshells with $j \le 7/2$ or even
$j = 9/2$ and are also applied in the present program. This is in contrast 
to most earlier atomic structure codes which are built on the cfp. 
For $j \le 7/2$, the rcfp have been taken from Rudzikas~\cite{R} while the
corresponding tables for $ j = 9/2 $ have been created by us using
Eq.\ (\ref{eq:second-b}) and the tabulations by Shalit and Talmi~\cite{ST} 
for the cfp. Similarly, a tabulation of the completely
reduced matrix elements of $W^{(k_q k_j)}$ have been obtained from the reduced
matrix elements of $T^{(k)}$ \cite{SSKR} and from the two relations 
(\ref{eq:unit-g}) and (\ref{eq:unit-h}) for the subshells
with $j \le 7/2$. Up to the present, the module \texttt{rabs\_{}rcfp} does 
not contain 
a full tabulation of the completely reduced matrix elements of 
$W^{(k_q k_j)}$ for $ j = 9/2 $ even though such an implementation might 
help considerably in the future in order to accelarate structure calculations 
on atoms having open $g_{\,9/2}$ and/or $h_{\,9/2}$ subshells. At present, 
these coefficients are calculated from Eq.\ (\ref{eq:unit-f})
each time they are needed. Also, the values of the cfp and the reduced
matrix elements of $T^{(k)}$ are calculated from Eqs.\ (\ref{eq:second-b}) or
(\ref{eq:unit-g}) and (\ref{eq:unit-h}), respectively.

\subsection{Interactive work}

The program RCFP is typically applied in its interactive mode.
In this mode, it replies immediately to the input as typed in by the user.
In the next section, we display several short dialogs for 
calculating individual coefficients and matrix elements. From the main 
menu of the RCFP component (see Fig.~1), 

\begin{figure}
\begin{small}
\begin{verbatim}
 RCFP: Calculation of coefficients of fractional parentage (cfp) and various 
  reduced matrix elements in jj-coupling (Fortran 90 version)
  (C) Copyright by G. Gaigalas and S. Fritzsche, Kassel (1999).
  
 Select one issue from the list for calculating:
  
   1:  coefficients of fractional parentage,
   2:  reduced coefficients of fractional parentage,
   3:  completely reduced matrix elements of the operator W^{k_q k_j},
   4:  reduced matrix elements of unit operator T^{(k)},
   b:  return to this menu,  
   q:  quit.
\end{verbatim}
\end{small}
{\bf Figure 1:}
\hspace{0.2cm}
{\rm The main menu of RCFP.}
\end{figure}

we need first to select the type of the quantity which is to be computed. 
For example, by entering \texttt{1} on the screen the user can calculate any
cfp in $jj$--coupling for subshells with $ j \le 9/2 $.
Similarly, a \texttt{2} supports the computation of rcfp,
and so on. Finally, a \texttt{q} will terminate the program.

\medskip

The input of the required quantum numbers needed for the computation of any
quantity is facilitated by the program. It is only necessary to type those
quantum numbers which cannot be derived 
automatically and which distinguish the individual coefficients and matrix 
elements. For calculating a cfp or a reduced matrix elements of $T^{(k)}$, 
for instance, the orbital quantum number $j$, the subshell occupation 
number $N$, the seniority quantum number $\nu$, and the subshell total 
angular momentum $J$ is needed in order to specify the bra--function uniquely.
Only if additional quantum numbers are indeed required for a unique
classification of the subshell states, the program will ask for the 
quantum number $Nr$.
A number of examples will illustrated the usage of RCFP below in section 4.

\subsection{Distribution and installation of the program}

RCFP has been developed as (a new) component of the RATIP package
\cite{Fritzsche/CFF/Dong:99}. To facilitate the combination with this package,
RCFP will be distributed as an archive file of the directory 
\texttt{ratip\_rcfp}. From this archive, first of all the file structure 
is reconstructed by the command \texttt{tar -xvf ratip\_rcfp.tar} on a 
UNIX workstation or any compatible environment. The directory 
\texttt{ratip\_rcfp} then contains the Fortran 90/95 module 
\texttt{rabs\_rcfp.f}, the (main) program \texttt{xrcfp.f} as well as the 
makefile \texttt{make-rcfp}. It also includes a number of examples in 
the subdirectory \texttt{test-rcfp} and a short \texttt{Read.me} which
explains further details about the installation. Since the same
file structure is preserved in both cases, the combination of RCFP with 
RATIP is simply achieved by running the command
\texttt{cp -r ratip\_rcfp/. ratip/.} --- Inside of the RATIP root directory, 
then
\texttt{make -f make-rcfp} will generate the executable \texttt{xrcf}, similarly
as for the other two components \texttt{xcesd99} \cite{Fritzsche/Anton:99}
and \texttt{xreos99} \cite{Fritzsche/CFF/Dong:99} of the RATIP package.
Like before, the name of the (Fortran 90/95) compiler and special compiler flags
can be overwritten in the header of the makefile. Although RCFP makes 
\textit{use} of four other modules which are part already of RATIP, no further
adaptation of the program is needed. At present, the RCFP program 
has been installed and tested under the operating systems Linux and AIX 
but, owing to the compliance of the Fortran 90/95 standard, no difficulties
should arise on any other platform.

\medskip

The subdirectory \texttt{test-rcfp} lists a number of examples 
which demonstrate the usage of the program. To each item in the main 
menu in Fig.\ 1,
a short file displays the full dialog to compute one or several individual
coefficients or (completely reduced) matrix elements. The file
\texttt{show-cfp-dialog}, for instance, reports the calculation of several
cfp for subshells with $ j = 9/2 $ including an example for
which the (additional) quantum number Nr need to be specified.

\medskip

Apart from the application of \texttt{rabs\_rcfp} in the RCFP program, this
module can be \textit{use}d also in other programs which, in the future, will
provide the angular coefficients for general matrix elements of one-- and
two--particle operators for $jj-$coupled functions.

\section{Examples}

To illustrate the use of RCFP in its interactive mode, we show three
examples concerning the calculation of rcfp and matrix
elements. We will just display the input (which has to be typed in by the user)
along with the given reply by the program. In order to support also an
occasional
usage of the program, the notation of the various coefficients and matrix
elements is kept as close as possible with their printed form [cf.\ section 2].
Moreover, all information which can automatically be deduced by the program is
simply provided by typing {\tiny \framebox{Enter}} at input time. For an 
improper
selection of quantum numbers  or any incomplete information, a short message is
printed explaining the failure before all previous (correct) input is repeated.
This saves the user from re-enter all of the previously typed
input just because of one single (mistyped) quantum number. 
In the following examples, we display the user's input in boldface
mode while the response of the program is shown in normal text mode.

\medskip

Our first examples displays the computation of the cfp
\newline
$\left( [7/2]^4, \, \nu =2, \, J=2 \, \{|\: [7/2]^3,\, \nu =3, \, J=3/2,\, 
\nu =1, \, j=7/2 \right)$; from the main menu in figure 1, we therefore select
the first item
\begin{small}
\begin{verbatim}
1
 Calculate a cfp (j^N  nu J {| j^{N-1} nu' J',  j) :
\end{verbatim}

\textbf{(7/2\^{}4 2 2} \texttt{{\tiny \framebox{Enter}}       \newline
  (7/2\^{}4 2 2 $\left\{| \right.$ 7/2 \^{}3}  \textbf{3 3/2,} 
\texttt{{\tiny \framebox{Enter}}                              \newline
  (7/2\^{}4 2 2 $\left\{| \right.$ 7/2 \^{}3 3 3/2,  1 7/2 ) =  2.53546276E-01
\newline
Continue
}

\end{small}
\medskip

Next, let us calculate the rcfp 
$\left( 9/2, \: \nu =5, \: J=1/2 \: ||| \: a^{(q j)} \: ||| \:
9/2, \: \nu =4, \: J=5 \right)$ 
for which we select item 2 from the main menu
\begin{small}
\begin{verbatim}
2
 Calculate a reduced cfp (j  nu J ||| a^{(1/2 j)} ||| j  nu' J') :
\end{verbatim}

\textbf{(9/2\ 5 1/2} \texttt{{\tiny \framebox{Enter}}       \newline
  (9/2\ 5 1/2 ||| a\^{}$\left\{ (1/2 \: j) \right\} |||$ 9/2} \textbf{4 5,} 
\texttt{{\tiny \framebox{Enter}}                              \newline
 Input must either start with symbol '(' or  end with symbol ')'; reenter ...
\newline
(9/2 5 1/2 ||| a\^{}$\left\{(1/2 \: j)\right\}$ ||| 9/2} \textbf{4 5)} 
\texttt{{\tiny \framebox{Enter}}                              \newline
(9/2 5 1/2 ||| a\^{}$\left\{(1/2 \: j)\right\}$ ||| 9/2 4 5) = 3.22490310E+00
\newline
Continue
}

\end{small}
\medskip

In our third example, finally, we ask for the value of one of the completely 
reduced matrix element of the $W^{(10)}$ operator, i.e.
$(j = 9/2, \: \nu = 4, \: J = 6, \: Nr= 2 \: ||| \: W^{ (1 0) } \: ||| \:
j = 9/2, \: \nu = 4, \: J = 6, \:  Nr= 2)$.
As mentioned in section 2, an additional quantum number $Nr$ is required for
a unique specification of the subshells states with $j=9/2$. Here, we start
by selecting item 3 from the main menu.
\begin{small}
\begin{verbatim}
3
 Calculate a completely reduced matrix element 
 (j   nu J ||| W^{k_q k_j} ||| j  nu' J') :
\end{verbatim}

\textbf{(9/2 4 6} \texttt{{\tiny \framebox{Enter}}              \newline
   Enter the additional state identifier Nr = 1 or 2. \newline
   (9/2 4 6 Nr=} \textbf{2} \texttt{{\tiny \framebox{Enter}}    \newline
   (9/2 4 6 Nr= 2 ||| W\^{}$ \left\{ \right. $} \textbf{ 1 0}  
\texttt{{\tiny \framebox{Enter}}                                \newline
   (9/2 4 6 Nr= 2 ||| W\^{}$ \left\{ \right. $ 1 0 
   $\left. \right\}$ ||| 9/2}  \textbf{4 6 )}                  
\texttt{{\tiny \framebox{Enter}}                                \newline
   (9/2 4 6 Nr= 2 ||| W\^{}$ \left\{ \right. $ 1 0 
   $\left. \right\}$ ||| 9/2 Nr=} \textbf{2 )}                 
\texttt{   {\tiny \framebox{Enter}}                             \newline
   (9/2 4 6 Nr= 2 ||| W\^{}$ \left\{ \right. $ 1 0 
   $\left. \right\}$ ||| 9/2 Nr= 2) = -3.94968353E+00       
\newline
Continue
}

\end{small}
\medskip

A very similar dialog occurs for the computation of any other coefficient
or reduced matrix element. In conclusion, RCFP has been developed as 
a new component of the RATIP package which enables the user to calculate 
standard quantities in the evaluation of many--electron matrix
elements explicitly. In the future, the underlying Fortran 90/95 module
\texttt{rabs\_{}rcfp} will be exploited also to calculate the Hamiltonian
matrix and further properties of free atoms from $jj-$coupled configuration
state functions.
The definition of the rcfp and the completely reduced matrix
elements and further improvements (see Gaigalas {\it et al}~\cite{method2}) 
will allow for faster
and more convinient computations than it is presently supported by standard
atomic structure programs. A module for calculating the angular coefficients 
for $jj-$coupled functions with respect to any (given)
scalar two--particle operator is currently under development.

\begin{small}

\end{small}
                 
\end{document}